\documentclass[pra,aps,twocolumn,superscriptaddress,showpacs,nofootinbib]{revtex4}
\usepackage{graphicx,graphics,epsfig}
\usepackage{amsmath}
\usepackage{verbatim}
\usepackage{color}
\usepackage[standard]{ntheorem}
\usepackage{subfigure}

\definecolor{myurlcolor}{rgb}{0,0,0.4}
\definecolor{mycitecolor}{rgb}{0,0.5,0}
\definecolor{myrefcolor}{rgb}{0.5,0,0}
\usepackage[pagebackref]{hyperref}
\hypersetup{colorlinks, linkcolor=myrefcolor,
citecolor=mycitecolor, urlcolor=myurlcolor}

\newcommand{\tr}{\mbox{tr}}
\def\endproof{\vrule height6pt width6pt depth0pt}

\newcommand{\ket}[1]{|#1\rangle}
\newcommand{\bra}[1]{\langle#1|}

\definecolor{nblue}{rgb}{0.2,0.2,0.8}
\definecolor{ngreen}{rgb}{0.2,0.8,0.2}
\definecolor{nred}{rgb}{0.8,0.2,0.2}
\definecolor{nblack}{rgb}{0,0,0}

\def\endproof{\vrule height6pt width6pt depth0pt}

\date{\today}

\begin{document}

\title{General tradeoff relations of quantum nonlocality in the Clauser-Horne-Shimony-Holt scenario}

\author{Hong-Yi Su}
\email{hongyisu@chonnam.ac.kr}
 \affiliation{Department of Physics Education, Chonnam National University, Gwangju 500-757, Republic of Korea}

\author{Jing-Ling Chen}
\affiliation{Theoretical Physics Division, Chern Institute of Mathematics, Nankai University,
 Tianjin 300071, People's Republic of China} \affiliation{ Centre for Quantum Technologies, National University of Singapore,
 3 Science Drive 2, Singapore 117543}

\author{Won-Young Hwang}
\email{wyhwang@jnu.ac.kr}
 \affiliation{Department of Physics Education, Chonnam National University, Gwangju 500-757, Republic of Korea}

\date{\today}

\begin{abstract}
General tradeoff relations present in nonlocal correlations of bipartite systems are studied, regardless of any specific quantum states and measuring directions. Extensions to multipartite scenarios are possible and very promising. Tsirelson's bound can be derived out in particular. The close connection with uncertainty relations is also presented and discussed.

\end{abstract}


\maketitle

\section{introduction}

Correlations of events have arguably been regarded as one of the most fundamental concepts in physics~\cite{bell}.
For multipartite scenarios, for instance, many attentions have been paid to entanglement, for which, e.g., concurrence of $\rho_{AB}$ and of $\rho_{AC}$ for an arbitrary tripartite state $\rho_{ABC}$ for Alice, Bob and Charlie satisfy the Coffman-Kundu-Wootters relation~\cite{CKW}, particularly implying a monogamous relation that Alice can only be maximally entangled with either Bob or Charlie, but not with both. Likewise, for Bell nonlocality~\cite{toner0}, which is like entanglement, party pairs $AB$ and $AC$ cannot share nonlocal correlations concurrently, but which is unlike entanglement, only in the two-setting scenario has this peculiar tradeoff been found to be present by far~\cite{CG04,ext}. It has been investigated that monogamous relations can also exist when it comes to contextual correlations and even a hybrid sort of correlations consisting of both contextuality and Bell nonlocality~\cite{dago}. The fact that quantum key distribution protocols based on nonlocality~\cite{QKD} can be proved secure can also be understood as the consequence of monogamous correlations among Alice, Bob, and Eve~\cite{QKD2}.

It is notable that of all the above studies more-than-three parties (or contexts) must be involved, in order to give a clear tradeoff in correlations~\cite{jens,costa}. On the other hand, the case of two parties, which serves as probably the simplest test bed for enormous \emph{gedanken} and utilitarian experiments, has nevertheless drawn relatively less attentions to researchers in this subject. Here, we shall demonstrate a general sort of two-party tradeoff relations of Bell nonlocality.

To that end let us first recall the tripartite case again.
It is known~\cite{toner0} that in this case a strict tradeoff can be found as $\langle I_{AB}\rangle^2+\langle I_{AC}\rangle^2\leq8$, where $I_{AB(C)}$ represents the Bell-Clauser-Horne-Shimony-Holt (Bell-CHSH) operator~\cite{chsh} for parties $A$ and $B$ ($C$), and $\langle I\rangle:=\tr\rho I$ denotes the expectation with $\rho$, such that $|\langle I_{AB(C)}\rangle|\leq2\sqrt{2}$ within quantum mechanics --- which is also known as the Tsirelson bound~\cite{tsirelson}.
The key features here are that the measuring directions of $A$ should be chosen to be the same in both terms, and that the forms of operators $I_{AB}$ and $I_{AC}$ are the same as well.
Then, a number of natural questions are immediately in order: If directions of $A$ differ in each term of the relation, or further, if the forms of $I_{AB}$ and $I_{AC}$ also differ with one another, does a similar tradeoff relation still exist? Despite partial answers~\cite{toner0}, one can see that even in the simple case of three parties such innocent questions may lead to complicated variations, which in the present paper we shall endeavour to avoid.

The form of our main results (see the theorem below) resembles that in the tripartite monogamy. However it is somewhat misleading to directly term it ``monogamy'' because the word is usually used to describe the peculiar multipartite relations, while our results apply to the bipartite cases. Essentially, the results can be seen as a witness of the quantum boundary upon general no-signaling theories, i.e., any violation will imply that the probabilities cannot be drawn from measurements upon quantum systems. In such a context, our inequalities are weaker than the so-called NPA hierarchy~\cite{npa} and the necessary and sufficient boundaries for the two-setting-two-outcome scenario~\cite{masanes}; but ours are comparably much easier to deal with and actually aim at a different task --- tradeoffs that may exist between Bell nonlocal correlations.

\section{the general framework of tradeoff relations}\label{framework}

Let us denote the set of measurements of all parties as $\vec M=(M_{A_1},M_{A_2},M_{B_1},M_{B_2},...)$,
and likewise the outcomes as $\vec X=(X_{A_1},X_{A_2},X_{B_1},X_{B_2},...)$.
For later convenience, we denote $\vec M_{ij\cdots}=(M_{A_i},M_{B_j},\cdots)$ and $\vec X_{ij\cdots}=(X_{A_i},X_{B_j},\cdots)$.

To study tradeoffs, one can have two options with clear physical meanings: either (I) to introduce a second set of measurements $\vec M'$, or (II) to introduce a second set of outcomes $\vec X'$, with $\vec M'\neq \vec M$ and $\vec X'\neq\vec X$. Option (I) means clearly that we are studying the relation between the correlations under $\vec M$ and under $\vec M'$.
Similarly, one can see option (II) as the relation between two sets of outcomes $\vec X$ and $\vec X'$. In fact, we can make a step further. Consider the two-dimensional case $\vec X_{ij\cdots}=\{0,1\}^{\otimes N}$. Due to $\sum_{\vec X_{ij\cdots}}P(\vec X_{ij\cdots}|\vec M_{ij\cdots})=1$, any other $P(\vec X_{ij\cdots}'|\vec M_{ij\cdots})$ can be transformed to a function of $P(\vec X_{ij\cdots}|\vec M_{ij\cdots})$, but in general the specific form of the inequality will change. Inspired from this, we would rather interpret option (II) as a generalized relation between two arbitrary forms of inequalities
under a certain $\vec M$.

Hence, according to the above formalism, it turns out that we are actually studying \emph{(I) the relation of a certain Bell inequality under various sets of measurements, or (II) that of various Bell inequalities under a certain set of measurements.}

A distinct feature here is that the number of parties is unnecessarily restricted to be over 3. For instance, one can consider only one party, though in this case we gain nothing more than Heisenberg's uncertainly relation for $P(\vec X_{i}|\vec M_{i})$ and $P(\vec X_{i}|\vec M_{i}')$ (along option (I)), or a simple result of probability theory: $P(\vec X_{i}|\vec M_{i})+P(\vec X_{i}'|\vec M_{i})\leq1$ (along option (II)). As shown below, highly nontrivial results arise when it comes to the two-party case, in which nonlocal correlations identified by Bell inequalities are taken into account.

Option (I) can be equivalent to (II) when we impose extra constraints on $\vec M'$. Sepcifically, \emph{(I) and (II) are equivalent to one another when $\vec M'$ is obtained from $\vec M$ by interchanging any pair of measurements for each party.} In a two-setting-two-outcome scenario where the measurement can be represented by directions $\{\hat a,\hat b,...\}$ for Alice, Bob, etc., for instance, one can interchange Alice's measurements, i.e., $\vec M'=(\hat a_2, \hat a_1, \hat b_1, \hat b_2...)$ with $\vec M=(\hat a_1, \hat a_2, \hat b_1, \hat b_2...)$, then this is exactly another way to say about two equivalent forms of, say CHSH inequality, under a certain $\vec M$: $\langle \mathcal{A}_1\mathcal{B}_1+\mathcal{A}_1\mathcal{B}_2+\mathcal{A}_2\mathcal{B}_1-\mathcal{A}_2\mathcal{B}_2\rangle$ and $\langle \mathcal{A}_2\mathcal{B}_1+\mathcal{A}_2\mathcal{B}_2+\mathcal{A}_1\mathcal{B}_1-\mathcal{A}_1\mathcal{B}_2\rangle$, which clearly belong to option (II).

With such a constraint on $\vec M'$ we are then able to explore both options at a time. Below, unless explicitly pointed out, we work along option (II) for the sake of technical simplicity.
But we must keep in mind that the same problem can of course be described, and resolved, along option (I) as well.

\section{Bell nonlocality in bipartite systems: main results}\label{main}


Let us first consider the inequivalent forms of the CHSH inequalities, obtained by interchanging the settings for each party (cf. equations listed in the theorem below). With a proper choice of measuring directions the state $\ket{\psi}=(\ket{00}+\ket{11})/\sqrt{2}$ violates the inequality maximally, i.e., $2\sqrt{2}>2$. Putting the same directions into any of its equivalent forms will result in a zero~\cite{zero}.

But in this example we have imposed two very strong requirements: (i) the state is pure and maximally entangled, and (ii) the measurements are chosen such that one of the inequality is maximally violated. In fact, both requirements can be discarded, while the tradeoff still holds. Then we have specifically

\noindent\textbf{Theorem:} Consider all the equivalent forms of the CHSH inequalities $|\langle I_\mu\rangle|\leq2$,
where $\langle I_\mu\rangle:=\tr\rho I_\mu$, $\mu=0,1,2,3$, along with (up to a global minus sign)
\begin{equation}\nonumber
\begin{split}
I_0&=+\mathcal{A}_1 \mathcal{B}_1+\mathcal{A}_1 \mathcal{B}_2+\mathcal{A}_2 \mathcal{B}_1-\mathcal{A}_2 \mathcal{B}_2,\\
I_1&=-\mathcal{A}_1 \mathcal{B}_1+\mathcal{A}_1 \mathcal{B}_2+\mathcal{A}_2 \mathcal{B}_1 +\mathcal{A}_2 \mathcal{B}_2,\\
I_2&=+\mathcal{A}_1 \mathcal{B}_1-\mathcal{A}_1 \mathcal{B}_2+\mathcal{A}_2 \mathcal{B}_1+\mathcal{A}_2 \mathcal{B}_2,\\
I_3&=+\mathcal{A}_1 \mathcal{B}_1+\mathcal{A}_1 \mathcal{B}_2-\mathcal{A}_2 \mathcal{B}_1+\mathcal{A}_2 \mathcal{B}_2,
\end{split}
\end{equation}
$\mathcal{A}_i, \mathcal{B}_j$ being Hermitian operators with eigenvalues $\pm1$. The following relations holds for any quantum state under an arbitrary set of projective measurements:
\begin{equation}\nonumber
\begin{split}
\langle I_\mu\rangle^2+\langle I_\nu\rangle^2\leq8, \;\;\forall\;\;\mu\neq\nu. \label{general}
\end{split}
\end{equation}

In other words, at most one of the CHSH inequalities can be violated. This is analogous to the usual monogamy relation of nonlocality of the tripartite case, but as discussed in the preceding section, we cannot term it monogamy in a usual sense, as we are considering the bipartite Alice-Bob case. Note also that Tsirelson's bound $2\sqrt{2}$ is an immediate result from the theorem.
In what follows we shall give a proof of the theorem.

\subsection{Proof of the theorem: setting up the stage}

Since the choice of labeling measuring settings is quite arbitrary, one can without loss of generality take the permutation-invariant form, $I_0$, as the starting point. To set up the stage, we opt to start by the qubit case of pure states and use the technique proposed in \cite{HHH} to rewrite the CHSH inequality as
\begin{equation}\nonumber
\begin{split}
\langle I_0\rangle
 =2\biggr[ (\hat a,T \hat c)\cos\theta+(\hat a',T \hat c')\sin\theta \biggr]\leq2.
 \end{split}
\end{equation}
Here $2\hat c\cos\theta=\hat b+\hat b'$, $2\hat c'\sin\theta=\hat b-\hat b'$, $(\cdot,\cdot)$ is the inner product in three-dimensional Euclidean spaces, and $T$ is a matrix with elements $t_{mn}=\tr\rho\;\sigma_m\otimes\sigma_n$, $m,n\in\{1,2,3\}$, where $\rho$ denotes an arbitrary two-qubit density matrix, $\sigma_m$'s are Pauli matrices, and $\hat a, \hat a'$ are measuring directions such that $\vec \sigma\cdot\hat a=\mathcal{A}_1, \vec \sigma\cdot\hat a'=\mathcal{A}_2$; similarly for $\mathcal{B}$. For the sake of convenience below, we further define $T\hat c=D\hat d$ and $T\hat c'=D'\hat d'$ with $\hat d$ and $\hat d'$ unit directions and $D$ and $D'$ taking real values.

\subsection{Proof for the case $\langle I_0\rangle-\langle I_1\rangle$}

\begin{figure}[tbp]
\includegraphics[width=70mm]{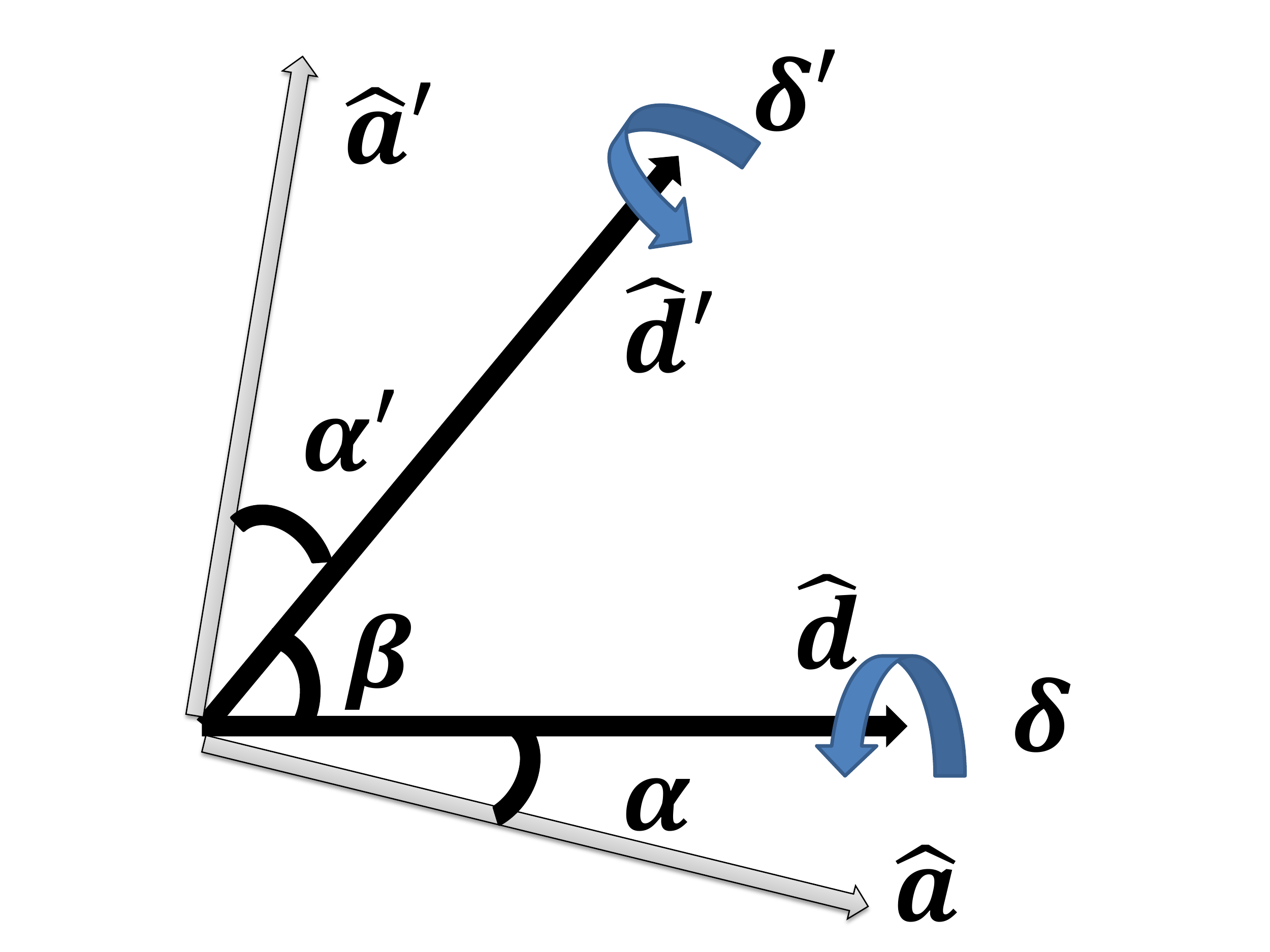}\\
\caption{The general measuring directions for the CHSH inequalities. Here $\hat a$ and $\hat a'$ denote directions for Alice corresponding to $\mathcal{A}_1$ and $\mathcal{A}_2$, respectively; $\hat d$ and $\hat d'$ denote directions for Bob defined above Eq.~(\ref{linear-eq1}).
 }\label{fig2}
\end{figure}

Let us consider $I_0$ and $I_1$:
\begin{eqnarray}
&\langle I_0\rangle=2\biggr[D(\hat a, \hat d)\cos\theta+D'(\hat a', \hat d')\sin\theta\biggr],&\label{linear-eq1}\\
&\langle I_1\rangle=2\biggr[D(\hat a', \hat d)\cos\theta-D'(\hat a, \hat d')\sin\theta\biggr],&\label{linear-eq2}
\end{eqnarray}
where the directions are shown in Fig.~\ref{fig2}. We introduce an extra parameter $\delta$ to represent the angle of rotation of $\hat a$ along $\hat d$
with their relative angle $\alpha$ unchanged; and similarly $\delta'$ for $\hat a'$ and $\hat d'$.

Let us see that with such a configuration the value of $\langle I_0\rangle$ does not depend on $\delta, \delta'$, explaining why the optimal set of directions to violate the CHSH inequality can always be found in a same plane. For problems involving, e.g., two equivalent inequalities as in the present paper, the rotation of $\hat a$ along $\hat d$ is nontrivial, however.

In general, let us assume
\begin{equation}\nonumber
\begin{split}
&(\hat a, \hat d)=\cos\alpha,\;\;(\hat a', \hat d')=\cos\alpha',\\
&(\hat a', \hat d)=\cos(\alpha'+\beta-\delta'):=v,\\
&(\hat a, \hat d')=\cos(\alpha+\beta-\delta):=u.
\end{split}
\end{equation}
According to simple solid geometry, we have $\min\{0,2\alpha\}\leq\delta\leq\max\{0,2\alpha\}$ and $\min\{0,2\alpha'\}\leq\delta'\leq\max\{0,2\alpha'\}$. This means that the inner products take smooth values between extreme cases of $\alpha+\beta$ and $\beta-\alpha$, and of $\alpha'+\beta$ and $\beta-\alpha'$, respectively --- that is, the last two inner products listed above must further satisfy
\begin{equation}\label{condition}
\begin{split}
&\min\{\cos(\alpha'+\beta),\cos(\beta-\alpha')\}\leq\\
&\;\;\;\;\;\;\;\;\;\;\;\;\;\;\;\;\;(\hat a', \hat d)\leq\max\{\cos(\alpha'+\beta),\cos(\beta-\alpha')\},\\
&\min\{\cos(\alpha+\beta),\cos(\beta-\alpha)\}\leq\\
&\;\;\;\;\;\;\;\;\;\;\;\;\;\;\;\;\;\;\;(\hat a, \hat d')\leq\max\{\cos(\alpha+\beta),\cos(\beta-\alpha)\}.\\
\end{split}
\end{equation}
These requirements on inner products are important to exclude unphysical solutions, because $a<b<c$ does not imply $\cos a<\cos b<\cos c$ in general.

Solve the linear equations (\ref{linear-eq1}) and (\ref{linear-eq2}) for  $D$ and $D'$ in terms of $\langle I_0\rangle$ and $\langle I_1\rangle$, we then achieve
\begin{equation}\nonumber
\begin{split}
D=\frac{1}{\cos\theta}\frac{\langle I_0\rangle\cos(\alpha+\beta)+\langle I_1\rangle\cos\alpha'}{\cos(2\alpha+\beta)+\cos(2\alpha'+\beta)+2\cos\beta},\\
D'=\frac{1}{\sin\theta}\frac{\langle I_0\rangle\cos(\alpha'+\beta)-\langle I_1\rangle\cos\alpha}{\cos(2\alpha+\beta)+\cos(2\alpha'+\beta)+2\cos\beta},
\end{split}
\end{equation}
which, due to $|D|^2\leq1$ and $|D'|^2\leq1$\footnote{The fact that $|D|^2$ and $|D'|^2$ are no greater than $1$ can be easily deduced for pure states as follows. Consider an arbitrary pure state in its Schmidt decomposition: $\ket{\phi}=\cos\vartheta\ket{00}+\sin\vartheta\ket{11}$. The corresponding $T$ matrix is in a diagonal form ${\rm diag}\{1,\sin2\vartheta,-\sin2\vartheta\}$, such that the square of any of its elements is no greater than $1$. Hence $|D|^2\leq1$ and $|D'|^2\leq1$.}, lead to the following relations
\begin{equation}\nonumber
\begin{split}
\biggr[\langle I_0\rangle &\cos(\alpha+\beta)+\langle I_1\rangle\cos\alpha'\biggr]^2\\
&\leq \cos^2\theta\biggr[\cos(2\alpha+\beta)+\cos(2\alpha'+\beta)+2\cos\beta\biggr]^2,\\
\biggr[\langle I_0\rangle&\cos(\alpha'+\beta)-\langle I_1\rangle\cos\alpha\biggr]^2\\
&\leq \sin^2\theta\biggr[\cos(2\alpha+\beta)+\cos(2\alpha'+\beta)+2\cos\beta\biggr]^2.
\end{split}
\end{equation}
Summing them up yields
\begin{eqnarray}
A \langle I_0\rangle^2+B \langle I_1\rangle^2+2C \langle I_0\rangle\langle I_1\rangle\leq r^2,\label{ellipse-g}
\end{eqnarray}
where
\begin{equation}\nonumber
\begin{split}
A=&u^2+v^2,\\
B=&\cos^2\alpha+\cos^2\alpha',\\
C=&u\cos\alpha'-v\cos\alpha,\\
r^2=&[\cos(2\alpha+\beta-\delta)+\cos(2\alpha'+\beta-\delta')\\
&+\cos(\beta-\delta)+\cos(\beta-\delta')]^2.
\end{split}
\end{equation}

Apparently, the left-hand side of (\ref{ellipse-g}) describes an ellipse. In order to cancel the cross term and write it more concisely, one can perform a rotation of the coordinate $(\langle{I}_0\rangle,\langle{I}_1\rangle)$ to $(\langle\mathcal{I}_0\rangle,\langle\mathcal{I}_1\rangle)$ by a rotation matrix, i.e.,
\begin{eqnarray}\nonumber
\left[\begin{matrix}
\langle\mathcal{I}_0\rangle\\
\langle\mathcal{I}_1\rangle
\end{matrix}\right]=
\left[
\begin{matrix}
\cos\xi & -\sin\xi\\
\sin\xi & \cos\xi
\end{matrix}\right]
\left[\begin{matrix}
\langle{I}_0\rangle\\
\langle{I}_1\rangle
\end{matrix}\right].
\end{eqnarray}
Then (\ref{ellipse-g}) becomes
\begin{equation}\nonumber
\begin{split}
A' \mathcal{I}_0^2+B' \mathcal{I}_1^2+2C' \mathcal{I}_0 \mathcal{I}_1\leq r^2,
\end{split}
\end{equation}
where
\begin{equation}\nonumber
\begin{split}
A'&=A\cos^2\xi+B\sin^2\xi-C\sin2\xi\\
&\;=[A+B+(A-B)\cos2\xi-2C\sin2\xi]/2,\\
B'&=A\sin^2\xi+B\cos^2\xi+C\sin2\xi\\
&\;=[A+B-(A-B)\cos2\xi+2C\sin2\xi]/2,\\
C'&=[(A-B)\sin2\xi+2C\cos2\xi]/2.
\end{split}
\end{equation}
By setting $C'=0$ one can achieve $2\xi+\eta=k\pi$ ($k=0,1,2...$)  where $\cos\eta=(A-B)/\sqrt{(A-B)^2+4C^2}$ and $\sin\eta=2C/\sqrt{(A-B)^2+4C^2}$.

Hence, we finally achieve two solutions to $A', B'$:\\
(i) $k$ is even:
\begin{equation}\nonumber
\begin{split}
A'&=\biggr[A+B+\sqrt{(A-B)^2+4C^2}\biggr]/2,\\
B'&=\biggr[A+B-\sqrt{(A-B)^2+4C^2}\biggr]/2,
\end{split}
\end{equation}
(ii) $k$ is odd:
\begin{equation}\nonumber
\begin{split}
A'&=\biggr[A+B-\frac{(A-B)^2-4C^2}{\sqrt{(A-B)^2+4C^2}}\biggr]/2,\\
B'&=\biggr[A+B+\frac{(A-B)^2-4C^2}{\sqrt{(A-B)^2+4C^2}}\biggr]/2.
\end{split}
\end{equation}

Therefore, for arbitrary directions the values of $\langle\mathcal{I}_0\rangle,\langle\mathcal{I}_1\rangle$ (or equivalently $\langle I_0\rangle, \langle I_1\rangle$) must fall within an ellipse-shaped region constrained by \begin{equation}\nonumber
\begin{split}
\frac{\langle\mathcal{I}_0\rangle^2}{\mathcal{U}^2}+\frac{\langle\mathcal{I}_1\rangle^2}{\mathcal{V}^2}\leq 1,
\end{split}
\end{equation}
with
\begin{equation}\nonumber
\begin{split}
\mathcal{U}^2=\frac{r^2}{A'}, \;\;\;\;\mathcal{V}^2=\frac{r^2}{B'}.
\end{split}
\end{equation}
When $k$ is even (odd), $\mathcal{U}$ ($\mathcal{V}$) is the minor (major) semi-axis. Hence, it is necessary to prove that $\mathcal{U}^2\leq8$ for odd $k$, and that $\mathcal{V}^2\leq8$ for even $k$. In fact, $\mathcal{V}^2$ for even $k$ is always no leass than $\mathcal{U}^2$ for odd $k$,  rendering the proof even more simplified --- that is, it is enough to prove the case of even $k$.

After a lengthy calculation for the case of even $k$, we obtain the major semi-axis
\begin{equation}\nonumber
\begin{split}
\mathcal{V}^2={r^2}/{B'}=8-2\Delta
\end{split}
\end{equation}
with $\Delta=L-R$ and
\begin{equation}\nonumber
\begin{split}
L=&2-u^2-v^2+\sin^2\alpha+\sin^2\alpha',\\
R=& \biggr[(u-\cos\alpha)^2+(v-\cos\alpha')^2\biggr]^{1/2}\\
&\times\biggr[(u+\cos\alpha)^2+(v+\cos\alpha')^2\biggr]^{1/2}.
\end{split}
\end{equation}
If the quantity $\Delta\geq0$ then our claimed result in the theorem can be proved. To show this is true, since both $L$ and $R$ are non-negative, we prove $\Delta':=L^2-R^2\geq0$ instead. Specifically, relation $\Delta'\geq0$ is equivalent to
\begin{eqnarray}
{A}''u^2+{B}''v^2+2{C}''uv\leq{R}^2\label{ellipse}
\end{eqnarray}
with
\begin{equation}\nonumber
\begin{split}
{A}''&=2-\cos^2\alpha,\\
{B}''&=2-\cos^2\alpha',\\
{C}''&=-\cos\alpha\cos\alpha',\\
{R}^2&=4-2\cos^2\alpha-2\cos^2\alpha'.
\end{split}
\end{equation}
Recalling that $u,v$ must satisfy Eqs.~(\ref{condition}), one can see that
the problem has now been reduced to that whether a rectangle region bounded by (\ref{condition})in the $uv$-plane falls entirely in the region bounded by the ellipse (\ref{ellipse}). In this problem, it suffices to verify only extremal points --- that is, if the four vertices of the rectangle fall in (\ref{ellipse}), then the entire region
spanned by their convex combinations will do also. It turns out that
\begin{equation}
\begin{split}
\Delta'&=\biggr[\cos[2(-1)^x\alpha+\beta]+\cos[2(-1)^y\alpha'+\beta]-2\cos\beta\biggr]^2\\
&{\rm for}\;\;u=\cos[(-1)^x\alpha+\beta],\;v=\cos[(-1)^y\alpha'+\beta],\\
&\;\;\forall x,y=0,1.\nonumber
\end{split}
\end{equation}
Hence, $\Delta\geq0$ holds for the entire physically allowed region of $u$ and $v$, thus leading to $\mathcal{V}^2\leq8$, with which we conclude that, for pure states, the relation $\langle I_0\rangle^2+\langle I_1\rangle^2\leq8$ holds under arbitrary measuring directions.

Let us now take account of mixed states with a simple geometric argument. Note that any mixed state is a convex combination of pure states, i.e., $\rho=\sum_i p_i \ket{\psi_i}\bra{\psi_i}$, where $\ket{\psi_i}$, without loss of generality, are chosen to be mutually orthogonal states. We define a vector $\vec I:=(\langle I_0\rangle,\langle I_1\rangle)$ in the $\langle I_0\rangle\langle I_1\rangle$-plane, then $\vec I^i=(\langle\psi_i| I_0|\psi_i\rangle,\langle\psi_i| I_1|\psi_i\rangle)$ in the plane represents a point that corresponds to the tradeoff relation with respect to $\ket{\psi_i}$. Due to the linearity of the trace operation the vector $\vec I(\rho)=(\sum_i p_i \langle\psi_i| I_0|\psi_i\rangle,\sum_i p_i \langle\psi_i| I_1|\psi_i\rangle)$ represents the tradeoff relation with respect to $\rho$. It is clear that such a vector must lie within the circle since this is the case for each $\vec I^i$. Therefore, the relation $\langle I_0\rangle^2+\langle I_1\rangle^2\leq8$ holds for any mixed states.

Thanks to results in~\cite{acin}, any dichotomic observable acting on a high-dimensional spaces can be decomposed as the one acting on no-more-than two-dimensional subspaces. Thus, the proof we give here also applies to arbitrary dimensional Hilbert spaces, provided that observables $A_i, B_j$ take dichotomic values $\pm1$.

\subsection{Proofs for the cases $\langle I_0\rangle-\langle I_2\rangle$ and $\langle I_0\rangle-\langle I_3\rangle$}


For these cases we can give a proof based on properties of Bell operators~\cite{scarani}. Because by definition $A_i^2=B_j^2=\openone$, we immediately have
\begin{equation}\nonumber
\begin{split}
I_0^2&-4\openone\otimes\openone=4\openone\otimes\openone-I_2^2=4\openone\otimes\openone-I_3^2\\
&=-[A_0,A_1]\otimes[B_0,B_1],
\end{split}
\end{equation}
yielding $I_0^2+I_2^2=I_0^2+I_3^2=8\openone\otimes\openone$. From relations
$\Delta^2 I_\mu=\langle I_\mu^2\rangle-\langle I_\mu\rangle^2\geq0$,
one finally achieves
$\langle I_0\rangle^2+\langle I_2\rangle^2\leq \langle I_0^2\rangle+\langle I_2^2\rangle \leq8$; similarly for $\langle I_0\rangle$ and $\langle I_3\rangle$. Combined with the proof for $\langle I_0\rangle$ and $\langle I_1\rangle$ in the previous section, we have already proved the results as claimed in the theorem. \hfill\endproof

\section{Comparison of numerical results with isotropic states}

We now consider for instance $I_0$ and $I_1$ with isotropic states
\begin{equation}\nonumber
\begin{split}
\ket{\Psi}&=V\ket{\phi}\bra{\phi}+(1-V)\frac{\openone\otimes\openone}{4},\\
\ket{\phi}&=\cos\vartheta\ket{00}+\sin\vartheta\ket{11},
\end{split}
\end{equation}
to verify our results in the theorem, and in the meantime to make a comparison if the requirement (ii), as stated in the paragraph above the theorem in Sec.~\ref{main}, is not discarded.

\begin{figure*}[tbp]
\subfigure[]{\includegraphics[width=40mm]{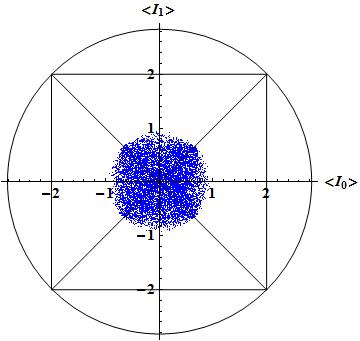}}\hspace{5mm}
\subfigure[]{\includegraphics[width=40mm]{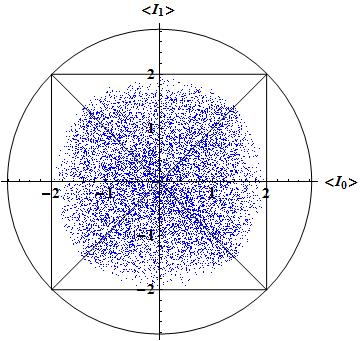}}\hspace{5mm}
\subfigure[]{\includegraphics[width=40mm]{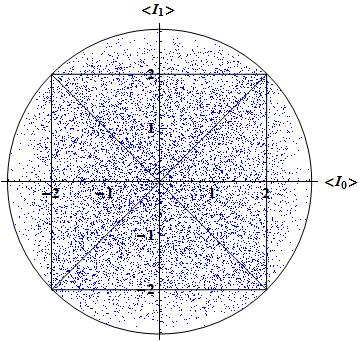}}\hspace{5mm}
\subfigure[]{\includegraphics[width=40mm]{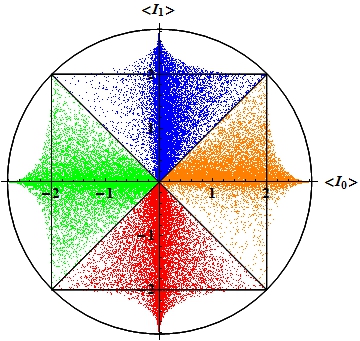}}
\caption{ (color online) The $\langle I_0\rangle \langle I_1\rangle$-plane for the isotropic states. Here the square inside the circle represents the region that admits local hidden variable models. For comparison we fix $\vartheta=\pi/4$, and take $V=\{1/3,1/\sqrt{2},1\}$ in subfigures (a), (b) and (c), respectively, with 50,000 sets of measuring directions randomly generated for each subfigure. In subfigure (d), the permitted region shrinks to the right (orange) quarter of the circle if it is further required that $\langle I_0\rangle$ be equal to its maximum under optimal directions; similarly, the permitted region to the left (green), top (blue) or bottom (red) quarter if $\langle I_0\rangle$ be minimum, $\langle I_1\rangle$ be maximum or $\langle I_1\rangle$ be minimum, respectively. In each quarter we randomly generate 50,000 states with respect to $V$ and $\vartheta$. }\label{fig5}
\end{figure*}

It can be seen from Fig.~\ref{fig5} that isotropic states distribute generally within the circle of radius $2\sqrt{2}$; however, the permitted region shrinks to an eight-pointed star-shaped one if we require that either $\langle I_0\rangle$ or $\langle I_1\rangle$ reaches its maximum or minimum. We have also considered maximally entangled mixed states as well as their convex combinations with product states. While we do not plot them here, they all similarly reproduce the eight-pointed star-shaped region if requirement (ii) is imposed.

\section{A viewpoint from uncertainty relations}

For maximally entangled pure states under optimal settings, the upper bound 8 in the theorem can be alternatively derived out from the uncertainty relation (3) or (4) proposed in \cite{pati}. From their formula, it is found that $\Delta^2I_0+\Delta^2I_1\geq8$,
which leads to $\langle I_0\rangle^2+\langle I_1\rangle^2\leq \langle I_0^2\rangle+\langle I_1^2\rangle-8\leq8$.
However, such a uncertainty-relation-based method cannot give $8$ under a set of arbitrary settings; examples can be found such that the upper bound given by the uncertainty relation could equal approximately $10.08$. Further, the effectiveness of the uncertainty relation in \cite{pati} becomes compromising when it comes to mixed states, e.g., the isotropic state, which is non-orthogonal with any quantum state. Our results in turn imply that at least in a four-dimensional system some tighter bound of uncertainty relations indeed exists (see also \cite{huang} where a variance-based method was used).

\section{Conclusion and outlooks}

In this paper, we have investigated the general tradeoff relations of Bell nonlocality. To demonstrate this particular tradeoffs specifically, we have taken the CHSH inequality and its equivalent forms into account and showed that at most one of the inequalities can be violated within quantum theory. The results hold for any quantum states with arbitrary sets of projective measurements. Tsirelson's bound can also be derived out of our results.

When no constraint is imposed and options (I) and (II) in Sec.~\ref{framework} are thus considered separately, we nevertheless believe that such a sort of tradeoff relations do not exist in general. For an instance along (I), if $\vec M'$ is chosen to be very close to $\vec M$, then both $I_\mu$ and $I_\nu$ could be close to being maximally violated --- no nontrivial tradeoff can be expected.
Whether under other circumstances, e.g., obtaining $\vec M'$ by rotating $\vec M$ with a finite angle, can a similar tradeoff still be found leaves us an open question.

The most surprising aspect to us in this paper is the fact that the bipartite system can exhibit so simple a tradeoff relation --- a circle --- it is meaningful as well to exploit whether such relations exist in multipartite systems, leaving us another open question no less worthy to explore as the first one.

In addition, one may wonder whether some more-than-three-setting Bell inequalities can be used here to derive tradeoff relations as well. The answer, we conjecture by referring to results in the usual multipartite monogamy, seems to be negative; however, further investigations are needed to confirm or disprove this conjecture. Also, more general measurements like the POVMs are worth being taken into account.

\acknowledgments

J.L.C. is supported by the National Basic Research Program (973 Program) of China under Grant No. 2012CB921900 and the Natural Science Foundations of China (Grants No. 11175089 and No. 11475089). This work is also supported by Institute for Information and Communications Technology Promotion (IITP) grant funded by the Korea Government (MSIP) (No. R0190-16-2028, Practical and Secure Quantum Key Distribution).


\begin{thebibliography}{99}
\bibitem{zero}
This is because under the optimal choice each correlator of the inequality is equal to $1/\sqrt{2}$, except the last one which is equal to $-1/\sqrt{2}$, so that a sum leads to $2\sqrt{2}$, whereas any change of the place of the minus sign in the sum makes the four correlators canceled out.
\end{thebibliography}
\end{document}